\documentstyle[floats,aps]{revtex}
\begin{document}
\renewcommand{\thefootnote}{\fnsymbol{footnote}}
\draft
\title{\large\bf 
  Algebraic Bethe Ansatz for Integrable Extended Hubbard Models Arising from
  Supersymmetric Group Solutions}

\author {  Anthony J. Bracken, 
 Xiang-Yu Ge,
 Mark D. Gould, Jon Links and
Huan-Qiang Zhou }

\address{Centre for Mathematical Physics, 
The University of Queensland,
		     Brisbane, Qld 4072, Australia}

\maketitle

\vspace{10pt}

\begin{abstract}
Integrable extended Hubbard models arising from
symmetric group solutions
are examined in the framework of the graded Quantum Inverse Scattering
Method.
The Bethe ansatz equations for all these models are derived by
using the algebraic Bethe ansatz method.
\end{abstract}

\pacs {PACS numbers: 71.20.Fd, 75.10.Jm, 75.10.Lp}



\def\a{\alpha}
\def\b{\beta}
\def\d{\delta}
\def\e{\epsilon}
\def\g{\gamma}
\def\k{\kappa}
\def\l{\lambda}
\def\o{\omega}
\def\t{\theta}
\def\s{\sigma}
\def\D{\Delta}
\def\L{\Lambda}


\def\beq{\begin{equation}}
\def\eeq{\end{equation}}
\def\bea{\begin{eqnarray}}
\def\eea{\end{eqnarray}}
\def\ba{\begin{array}}
\def\ea{\end{array}}
\def\no{\nonumber}
\def\le{\langle}
\def\re{\rangle}
\def\lt{\left}
\def\rt{\right}

\vskip.3in

\section{Introduction\label{int}}

The study of exactly solvable models of strongly correlated electrons
is important for understanding fundamental aspects of statistical mechanics.
It is relevant to many realistic physical systems such as
high-$T_c$ superconductors.
The quantum
inverse scattering method (QISM) has been applied to solve various
strongly correlated electron models \cite{EK94}.
Two models which have attracted a great deal of attention
are the Hubbard model,
which is derived from
an $R$-matrix with non-additive spectral parameter,
and which was investigated  
by Shastry \cite{Sh},
and its strong coupling limit the $t-J$ model.
Essler and Korepin established
the integrability of the $t-J$ model by obtaining an infinite set of
conserved
quantities, and studied the model in the framework of the graded QISM
\cite{EK}.
A integrable anisotropic ($q$-deformed) supersymmetric $t-J$ model
was proposed by
Foerster and Karowski \cite{FK}. Woynarovich \cite{Wo} applied
the finite-size correction of the ground state
energy to obtain the low-lying gapless excitation spectrum around the
ground state. Frahm and Korepin \cite{FKo}
obtained the critical exponents for
various correlation functions.
Furthermore, the Bethe ansatz solution and
conformal properties of the $q$-deformed $t-J$ model
were studyed by Bariev, Kl\"umper, Schadschneider and Zittartz \cite{BKSZ}.

By considering other representations of quantum superalgebras,
new integrable, strongly correlated electron models of interest,
such as
the supersymmetric Essler-Korepin-Schoutens (EKS)
extended Hubbard model \cite{EKS}, have been
proposed. This $gl(2|2)$ supersymmetric model contains the
supersymmetric $t-J$ model as a submodel and can be interpreted as the
Hubbard model plus moderate nearest-neighbour interactions. The complete
solution
of the EKS model by the algebraic Bethe ansatz has
been obtained \cite{EKS1}. The mathematical issue of the completeness of
the solution has been settled \cite{Sch}, and the physics content of the
solution, low lying excitations in particular, has been studied 
\cite{EK1}.
Another non-standard
integrable model of strongly correlated electrons is the
Bariev chain \cite{Ba},
which has
an $R$-matrix with non-additive spectral parameter 
as investigated by Zhou \cite{Zh}.
A generalization of the Hubbard model
 with Lie superalgebra $gl(2|1)$ supersymmetry,
 the supersymmetric $U$ model,
was discovered in \cite{BGLZ}, and has been
investigated by several groups
\cite{BF1,PF1,RM,HGL}.
 An extension of this model, a $q$-deformed version
 with quantum superalgebra $U_q(gl(2|1))$
 supersymmetry, was also proposed \cite{BKZ,GHLZ}.
Thermodynamic properties of the EKS model and the supersymmetric $U$
model have been studied using Wiener-Hopf techniques and the critical
exponents of correlation functions by using methods of conformal field
theory \cite{BKZ}.
Recently, the eight-state
supersymmetric $U$ model
of strongly correlated electrons
with the Lie superalgebra $gl(3|1)$ symmetry,
and the two-parameter ($q$-deformed) eight-state supersymmetric fermion
model with quantum superalgebra $U_q(gl(3|1))$ symmetry,
were introduced \cite{GZZ,GGZZ}.

Dolcini and Montorsi \cite{DM} introduced  
integrable extended Hubbard Hamiltonians from symmetric
group solutions. 
One of the aims of this work is to show the solution of
these models via the QISM.
The most general form of the extended Hubbard model 
invariant under spin-flip and conserving the total number of electrons
and magnetization, first considered in \cite{BKS}, is described by
the Hamiltonian 
\bea
H&=&\mu_e \sum _{j,\s}n_{j,\s} -\sum_{<j,k>,\s}[t-X(n_{j,-\s}+n_{k,-\s})
+\tilde{X}n_{j,-\s}n_{k,-\s}]c_{j,\s}^\dagger c_{k,\s}\no\\
& &+U \sum _j n_{j,\uparrow}n_{j,\downarrow} + \frac{V}{2}\sum
_{<j,k>}n_jn_k +\frac{W}{2} \sum _{<j,k>,\s,\s'} 
c_{j,\s}^\dagger c_{k,\s'}^\dagger c_{j,\s'}c_{k,\s}\no\\
& & +Y \sum _{<j,k>} c_{j,\uparrow}^\dagger c_{j,\downarrow}^\dagger
  c_{k,\downarrow} c_{k,\uparrow}
+P \sum _{<j,k>} n_{j,\uparrow}n_{j,\downarrow}n_k
+\frac{Q}{2} \sum _{<j,k>} n_{j,\uparrow}n_{j,\downarrow}
n_{k,\uparrow}n_{k,\downarrow},
\label{ham}
\eea
where $\mu_e$ is the chemical potential.
Here, electrons on a lattice are described by
canonical Ferimi operators
$c_{j,\a}$ and
$c_{j,\a}^\dagger$ satisfying the anticommutation relations given by
$\{c_{i,\a}^\dagger, c_{j,\b}\}=\d_{ij}\d_{\a\b}$, where $i,j,=1,2,..., $
and $\a,\b=\uparrow,\downarrow$.
Also $n_{j,\a}=c_{j,\a}^\dagger c_{j,\a},n_j=n_{j,\downarrow}+n_{j,\uparrow}$.
In (\ref{ham}) the term $t$ represents the band energy of the electrons,
while
the subsequent terms describe their Coulomb interaction energy in a narrow
band approximation: $U$ parametrizes the on-site diagonal interaction,
$V$ the neighboring site charge interaction, $X$ the bond-charge
interaction, $W$ the exchange term, and
$Y$ the pair-hopping term. Moreover, additional many-body coupling terms
have been included in agreement with \cite{BKS} : $\tilde{X}$ correlates
hopping with on-site occupation number, and $P$ and $Q$ describe three-
and four-electron interactions. 

As was shown in \cite{DM}, the integrability of the model lies in 
the fact that
there exists a  solution of the   Yang-Baxter
equation which takes the form
$$
\check{R}(u)= 1+ u \Pi
$$
where $u$ is the spectral parameter and 
\small
\beq
\Pi=
\left ( \begin {array} {cccccccccccccccc}
\s^d_{11}&0&0&0&0&0&0&0&0&0&0&0&0&0&0&0\\
0&\s^d_{12}&0&0&\s^o_{12}&0&0&0&0&0&0&0&0&0&0&0\\
0&0&\s^d_{13}&0&0&0&0&0&\s^o_{13}&0&0&0&0&0&0&0\\
0&0&0&\s^d_{14}&0&0&0&0&0&0&0&0&\s^o_{14}&0&0&0\\
0&\s^o_{12}&0&0&\s^d_{21}&0&0&0&0&0&0&0&0&0&0&0\\
0&0&0&0&0&\s^d_{22}&0&0&0&0&0&0&0&0&0&0\\
0&0&0&0&0&0&\s^d_{23}&0&0&\s^o_{23}&0&0&0&0&0&0\\
0&0&0&0&0&0&0&\s^d_{24}&0&0&0&0&0&\s^o_{24}&0&0\\
0&0&\s^o_{13}&0&0&0&0&0&\s^d_{31}&0&0&0&0&0&0&0\\
0&0&0&0&0&0&\s^o_{23}&0&0&\s^d_{32}&0&0&0&0&0&0\\
0&0&0&0&0&0&0&0&0&0&\s^d_{33}&0&0&0&0&0\\
0&0&0&0&0&0&0&0&0&0&0&\s^d_{34}&0&0&\s^o_{34}&0\\
0&0&0&\s^o_{14}&0&0&0&0&0&0&0&0&\s^d_{41}&0&0&0\\
0&0&0&0&0&0&0&\s^o_{24}&0&0&0&0&0&\s^d_{42}&0&0\\
0&0&0&0&0&0&0&0&0&0&0&\s^o_{34}&0&0&\s^d_{43}&0\\
0&0&0&0&0&0&0&0&0&0&0&0&0&0&0&\s^d_{44}
\end {array}  \right ),
\label{Pi}
\eeq
\normalsize  
Indeed, after making the assignment
\bea
|1\re\,=|0\re\,,~~~
  |2\re\,={|\downarrow\uparrow\re\,}_j
 = c_{j,\downarrow}^\dagger c_{j,\uparrow}^\dagger|0\re\,,~~~
|3\re\,={|\uparrow\re\,}_j
=c_{j,\uparrow}^\dagger|0\re\,,~~~
  |4\re\,={|\downarrow\re\,}_j
  =c_{j,\downarrow}^\dagger|0\re\,,~~~
\label{con}
\eea
one can show that $H$ can be expressed as the sum over a graded
{\em generalized permutator},
\beq
H=\sum^L_{j=1}\Pi_{j,j+1}
\label{HPi}
\eeq
where the operator $\Pi_{j,j+1}$ permutes the four possible
configurations (\ref{con}) between the sites $j$ and $j+1$; viz. 
$$
\Pi=\s^d_{ik}(e_{ii}\otimes_s e_{kk}) + 
\s^o_{ik}(e_{ik}\otimes_s e_{ki})
$$
where
$\s^d_{ik}(e_{ii}\otimes e_{kk})$ are  diagonal terms and
$\s^o_{ik}(e_{ik}\otimes e_{ki})$ are  off-diagonal terms.
At this point we would like to mention that the construction here bears
some resemblance to those by Alcaraz and Bariev \cite{AB} and Maassarani
\cite{M} for solutions of the Yang-Baxter equation based on
representations of the Hecke algebra. 

It is clear that this form of interaction conserves the individual
numbers $N_{\uparrow}$ and $N_{\downarrow}$ of electrons with spin up
and spin down respectively, and the numbers $N_l$ and $N_h$ of doubly occupied
(local electron pairs) and empty sites (holes). We will choose the
following conventions throughout this paper:

$N_{\uparrow}$=number of single electrons with spin up

$N_{\downarrow}$=number of single electrons with spin down

$N_e$=$N_{\uparrow}+N_{\downarrow}$=number of single electrons 

$N_l$=number of local electron pairs 

$N_h$=number of holes 

$N_b$=$N_h+ N_l$=number of ``bosons''. 

To be specific, we give the relations between $\s^d_{i,k}$, 
$\s^o_{i,k}$ and parameters in the
Hamiltonian (\ref{ham}): 
\bea
&&\s^d_{11}=c,~~~ \s^d_{22}=c+Q+U+4P+2\mu_e+4V-2W,~~~ \s^d_{33}=\s^d_{44}=
c+V-W+\mu_e\no\\
&&\s^d_{12}=\s^d_{21}=c+\mu_e +\frac{U}{2},~~~ \s^d_{13}=\s^d_{14}
=\s^d_{31}=\s^d_{41}=c+\frac{\mu_e}{2},\no\\
&&\s^d_{23}=\s^d_{24}
=\s^d_{32}=\s^d_{42}=c+P+\frac{3\mu_e}{2}+\frac{U}{2}+2V-W,~~~
\s^d_{34}=\s^d_{43}=c+V +\mu_e\no\\
&&\s^o_{12}=Y,~~~\s^o_{13}=\s^o_{14}=-t,~~~\s^o_{23}=\s^o_{24}=-t+\tilde{X},
~\s^o_{34}=-W.
\no \eea

It turns out that actually there are 96 different possible choices of
values of the physical parameters in (1). They can be cast into six
groups as (\ref{Pi}) shows.
\small
\bea
\begin{tabular}{c||c||c||c||c||c||c||}
&$H_1(s_1,...,s_5)$&$H_2(s_1,...,s_5)$&$H_3(s_1,s_2,s_3)$&$H_4(s_1,s_2,s_3)$&$H_5(s_1,s_2,s_3)$ &$H_6(s_1,s_2,s_3)$\\
$t$&1&1&1&1&1&0\\
$X$&1&1&1&1&1&0\\
$\tilde{X}$&$1+s_2$&$1+s_2$&$1+s_2$&$1+s_2$&1&1\\
$U$&$2s_1$&$2s_1$&$4s_1$&$4s_1$&$2s_1$&$-2s_1$\\
$V$&$s_1$&$s_1+s_4$&$s_1$&$s_1+s_3$&$s_1+s_3$&0\\
$W$&$s_4$&0&$s_3$&0&0&0\\
$Y$&$s_3$&$s_3$&0&0&$s_2$&$s_2$\\
$P$&$s_4-s_1$&$-s_1-2s_4$&$s_3-2s_1$&$-2(s_1+s_3)$&$-(s_1+s_3)$&0\\
$Q$&$-2s_4+s_1+s_5$&$4s_4+s_1+s_5$&$4s_1-2s_3$&$4(s_1+s_3)$&$s_1+s_3$&$s_1+s_3$\\
$\mu_e$&$-2s_1$&$-2s_1$&$-2s_1$&$-2s_1$&$-2s_1$&0\\
$c$&$s_1$&$s_1$&$s_1$&$s_1$&$s_1$&$s_1$
\end{tabular}
\no \eea
\normalsize
Above, there is the  restriction $s_i=\pm 1,\, i=1,...,5.$ 

We now construct the eigenstates of the Hamiltonians of the
one-dimensional model in the above six groups, using the QISM. 
We use the $R$ matrix
satisfying the Yang-Baxter equation and introduce an $L$ operator
constructed directly from the $R$ matrix of the twisted representation.
The quantum Yang-Baxter equation can be written as the operator
equation:
\beq
\check{R}(\l-\mu)L_j(\l)\otimes L_j(\mu)
=L_j(\mu)\otimes L_j(\l)\check{R}(\l-\mu)
.\label {qybe}
\eeq
Here $\otimes$ denotes the graded tensor product defined by
$$
(A \otimes B)_{ij,kl}=(-1)^{(\e_i+\e_j)\e_k}A_{ij}B_{kl}
$$
where $\e_i\in {\bf Z}_2$ denotes the grading of the index $i$. 
We chose to adopt the bosonic, bosonic, fermionic and fermionic (BBFF)
grading 
$\e_1=\e_2=0,\,\e_3=\e_4=1$ on the indices labelling the basis vectors.

We now proceed to establish the relation between the Hamiltonian  (\ref{ham})
and the transfer matrix $\tau (\l)$, which is the supertrace of the
monodromy matrix $T(\l)$ defined by
$$
T(\l)=L_L(\l) L_{L-1}\cdots L_1(\l).
$$ From (\ref{qybe}) it follows that
\bea
\check{R}(\l-\mu)T(\l)\otimes T(\mu)
=T(\mu)\otimes T(\l) \check{R}(\l-\mu).
\label{YBA}
\eea
Thus we have 
$$
[\tau (\l), \tau (\mu)]=0
$$
and so the $\tau(\l)$ form a one-parameter family of commuting operators. The
transfer matrices may be taken as integrals of the motion, and so we 
obtain an infinite number of higher conservation laws of the model.

\section{algebraic Bethe ansatz for group 1}

Having recalled the quantum integrability of the models, let us
use the nested
algebraic Bethe ansatz method to find the eigenvalues of the transfer
matrices. As will be shown, the Bethe ansatz solutions for the six
different groups take on differing forms. In particular, the number of
levels of Bethe ansatz nestings ranges from 0 to 2. Moreover, some cases
do not admit a unique reference state. In such an instance, we are
forced to use a subspace of reference states to perform the
calculations. This type of procedure was first investigated by Abad and
R\'{\i}os \cite{AR} for the case of alternating $su(3)$ representations and
we will adopt this formalism where necessary.

We start from the first group
with BBFF
grading. The explicit form of the $R$-matrix is
\small
\bea
&&\check{R}(u)=1+u\Pi=\no\\
&&\left ( \begin {array} {cccccccccccccccc}
1+s_1u&0&0&0&0&0&0&0&0&0&0&0&0&0&0&0\\
0&1&0&0&s_3u&0&0&0&0&0&0&0&0&0&0&0\\
0&0&1&0&0&0&0&0&-u&0&0&0&0&0&0&0\\
0&0&0&1&0&0&0&0&0&0&0&0&-u&0&0&0\\
0&s_3u&0&0&1&0&0&0&0&0&0&0&0&0&0&0\\
0&0&0&0&0&1+s_5u&0&0&0&0&0&0&0&0&0&0\\
0&0&0&0&0&0&1&0&0&s_2u&0&0&0&0&0&0\\
0&0&0&0&0&0&0&1&0&0&0&0&0&s_2u&0&0\\
0&0&-u&0&0&0&0&0&1&0&0&0&0&0&0&0\\
0&0&0&0&0&0&s_2u&0&0&1&0&0&0&0&0&0\\
0&0&0&0&0&0&0&0&0&0&1-s_4u&0&0&0&0&0\\
0&0&0&0&0&0&0&0&0&0&0&1&0&0&-s_4u&0\\
0&0&0&-u&0&0&0&0&0&0&0&0&1&0&0&0\\
0&0&0&0&0&0&0&s_2u&0&0&0&0&0&1&0&0\\
0&0&0&0&0&0&0&0&0&0&0&-s_4u&0&0&1&0\\
0&0&0&0&0&0&0&0&0&0&0&0&0&0&0&1-s_4u
\end {array}  \right ),
\label{R1}
\no \eea
\normalsize
To utilize the  framework of the QISM, we write down the $L$ operator
\beq
{L}_j(u)=\frac{1}{1+s_1u}P\check{R}(u),
\label{lpr}
\eeq
where $P$ is the graded permutation operator,
\beq
P=\sum_{ij}(-1)^{[j]} e_{ij}\otimes e_{ji}.
\label{P}
\eeq
We choose the local vacuum state as $|0\re\,_j=(0,0,0,1)^t$. Acting
the $L$-operator on this local vacuum state, we have
\beq
{L}_j(u){|0\re\,}_j=
 \left ( \begin {array}
{cccc}
 1&*&*&*\\
 0& s_3a(u)&0&0\\
 0&0&-a(u)&0\\ 
 0&0&0&-a(u)
     \end {array} \right ){|0\re\,}_j,
\label{u1}
\eeq
with $a(u)=u/(1+s_1u)$. 
Define the vacuum state as $|0\re\,=\otimes^L_{j=1}{|0\re\,}_j$. Using
the standard QISM,
we represent the monodromy matrix in the
following way:
\beq
{T}(u)={L}_L(u){L}_{L-1}(u)\cdots {L}_1(u) \equiv
 \left ( \begin {array}
{cccc}
 A(u)&B_1(u)& B_2(u)&B_3(u)\\
 C_1(u)& D_{11}(u)&D_{12}(u)&D_{13}(u)\\
 C_2(u)&D_{21} (u)& D_{22}(u)&D_{23}(u)\\
 C_3(u)&D_{31} (u)& D_{32}(u)&D_{33}(u)
     \end {array} \right ),
\label{T11}
\eeq
The transfer matrix with periodic boundary conditions is thus written
explicitly as 
\beq
\tau(u)=A(u)+D_{11}(u)-D_{22}(u)-D_{33}(u).
\label{tm}\eeq
The action of the monodromy matrix on the pseudo-vacuum state is 
\bea
A(u) {|0\re\,} &=& 
 {|0\re\,} ~~~
D_{11}(u) {|0\re\,}= [s_3a(u)]^L
 {|0\re\,} ,~~~
D_{22}(u) {|0\re\,} = 
D_{33}(u) {|0\re\,}= [-a(u)]^L
 {|0\re\,},\no\\
B_k(u) {|0\re\,} & \neq& 0,~~~
C_k(u) {|0\re\,} =  0,~~~
D_{kl}(u) {|0\re\,} =  0~~~~~~~( k \neq l,~~k,l=1,2,3).
\label{T111}
\eea
Substituting (\ref{T11}) into the Yang-Baxter algebra (\ref {YBA}) 
we may deduce the following commutation relations:
\bea
D_{ab}(\mu)B_c(\l) &=&
S_a[(-1)^{\e_a\e_p}\frac{r(\mu-\l)^{dc}_{pb}}{a(\mu-\l)}B_p(\l)D_{ad}(\mu)
-(-1)^{\e_a\e_b}\frac{b(\mu-\l)}{a(\mu-\l)}B_b(\mu)D_{ac}(\l)],\no\\
A(\mu)B_c(\l) &=&
S_c[\frac{1}{a(\l-\mu)}B_c(\l)A(\mu)-\frac{b(\l-\mu)}{a(\l-\mu)}B_c(\mu)A(\l)],
\no\\
B_{a_1}(\l)B_{a_2}(\mu) &=&
r(\l-\mu)^{b_1a_2}_{b_2a_1}B_{b_2}(\mu)B_{b_1}(\l),
\no \eea
where
\bea
r(u)&=&b(u){I^{(1)}}^{ab}_{cd} +a(u){\Pi^{(1)}}^{ab}_{cd}\no\\
a(u)&=&\frac{u}{1+s_1u},~~~b(u)=\frac{1}{1+s_1u}.
\no \eea
Here ${\Pi^{(1)}}^{ab}_{cd}$ is the $9\times9$ submatrix
of $\Pi$,
\bea
{\Pi}^{(1)}=
\left ( \begin {array} {ccccccccccc}
s_5&0&0&0&0&0&0&0&0\\
0&0&0&s_2&0&0&0&0&0\\
0&0&0&0&0&0&s_2&0&0\\
0&s_2&0&0&0&0&0&0&0\\
0&0&0&0&-s_4&0&0&0&0\\
0&0&0&0&0&0&0&-s_4&0\\
0&0&s_2&0&0&0&0&0&0\\
0&0&0&0&0&-s_4&0&0&0\\
0&0&0&0&0&0&0&0&s_4
\end {array}  \right ),
\no
\eea
corresponding to the grading $\e_1=0, \e_2=\e_3=1$ and $S_1=\frac{1}{s_3},
S_2=-1, S_3=-1$. We assume the eigenvectors of the transfer matrix to
be
$$
B_{a_1}(\l_1)B_{a_2}(\l_2)\cdots B_{a_n}(\l_n)|0\re\,F^{a_n \cdots a_1}
$$
where $F^{a_1 \cdots a_n}$ is a function of the spectral parameters
$\l_j$. Acting the transfer matrix (\ref{tm}) on the above vector, we have
\bea
&&\tau(u)B_{a_1}(\l_1)B_{a_2}(\l_2)\cdots B_{a_n}(\l_n)|0\re\,
F^{a_n \cdots a_1}\no\\
&=&[A(u)+D_{11}(u)-D_{22}(u)-D_{33}(u)]
B_{a_1}(\l_1)B_{a_2}(\l_2)\cdots B_{a_n}(\l_n)|0\re\,F^{a_n \cdots a_1}\no\\
&=&s_3^{-N_l}(-1)^{N_e}\prod^n_{j=1}\frac{1}{a(\l_j-u)}
B_{a_1}(\l_1)B_{a_2}(\l_2)\cdots B_{a_n}(\l_n)|0\re\,F^{a_n \cdots a_1}\no\\
&&+[a(u)]^Ls_3^{L-n}(-1)^{N_e+N_{\downarrow}}
\prod^n_{j=1}\frac{1}{a(\l_j-u)}\tau^{(1)}(u,\{\l_k\})^{a_n \cdots
a_1}_{b_n \cdots b_1}
B_{b_1}(\l_1)B_{b_2}(\l_2)\cdots B_{b_n}(\l_n)|0\re\,F^{a_n \cdots a_1}\no\\
&&+u.t.
\no \eea
where $u.t.$ denotes unwanted terms, and $\tau^{(1)}(u)$ is the nested
transfer martix. Denote the eigenvalues of $\tau(u)$ and $\tau^{(1)}(u)$
by $\Lambda(u)$ and $\Lambda^{(1)}(u)$. We have
$$
\Lambda(u)
=s_3^{-N_l}(-1)^{N_e}\prod^n_{j=1}\frac{1}{a(\l_j-u)}
+[a(u)]^Ls_3^{L-n}(-1)^{N_e+N_{\downarrow}}
\prod^n_{j=1}\frac{1}{a(\l_j-u)}\Lambda^{(1)}(u).
$$
In order to cancel the unwanted terms, we need the following Bethe
ansatz equations
$$
\Lambda^{(1)}(\l_k)
=s_3^{-N_l}(-1)^{N_e}
[s_3^{L-n}(-1)^{N_e+N_{\downarrow}}]^{-1}
[a(\l_k)]^{(-L)}
\prod^n_{\stackrel{j=1}{j \neq k}}\frac{a(\l_k-\l_j)}{a(\l_j-\l_k)}.
$$
The nested transfer matrix is written as the supertrace on the auxiliary
space for the reduced monodromy matrix which satisfies the Yang-Baxter
relation.
$$
\tau^{(1)}(u,\{\l_k\})={\rm str[diag}(\frac{1}{s_3},-1,-1){L}^{(1)}_n(u-\l_n)
{L}^{(1)}_{n-1}(u-\l_{n-1}) \cdots {L}^{(1)}_1(u-\l_1)]
$$
\bea
r(\l-\mu){T}^{(1)}_n(\l)\otimes {T}^{(1)}_n(\mu)
={T}^{(1)}_n(\mu)\otimes {T}^{(1)}_n(\l) r(\l-\mu).
\label{YBAr}
\eea
If we write
\beq
{T}^{(1)}_n(u)={L}^{(1)}_n(u){L}^{(1)}_{n-1}(u)\cdots
{L}^{(1)}_1(u) \equiv
 \left ( \begin {array}
{ccc}
 A^{(1)}(u)&B^{(1)}_1(u)& B^{(1)}_2(u)\\
 C^{(1)}_1(u)& D^{(1)}_{11}(u)&D^{(1)}_{12}(u)\\
 C^{(1)}_2(u)& D^{(1)}_{21}(u)&D^{(1)}_{22}(u)
     \end {array} \right ),
\label{T21}
\eeq
the $L^{(1)}$ operator is 
$$
{L}^{(1)}_j(u)=P^{(1)}r_j(u),
$$
where $P^{(1)}$ is the $9\times9$ permutation operator
\beq
P^{(1)}=\sum_{ij}(-1)^{\e_j} e_{ij}\otimes e_{ji}, 
\label{P1}
\eeq
corresponding to the
grading $\e_1=0,\e_3=\e_3=1$. Now (\ref{YBAr}) and $r(u)$ imply that
\bea
D^{(1)}_{ab}(\mu)B^{(1)}_c(\l) &=&
\frac{1}{s_2}[
(-1)^{\e_a\e_p}\frac{r^{(1)}(\mu-\l)^{dc}_{pb}}{a^{(1)}(\mu-\l)}
B^{(1)}_p(\l)D^{(1)}_{ad}(\mu)
-(-1)^{\e_a\e_b}\frac{b^{(1)}(\mu-\l)}{a^{(1)}(\mu-\l)}B_b(\mu)D_{ac}(\l)],\no\\
A^{(1)}(\mu)B^{(1)}_c(\l) &=&\frac{1}{s_2}[
\frac{1}{a^{(1)}(\l-\mu)}B^{(1)}_c(\l)A^{(1)}(\mu)-\frac{b^{(1)}(\l-\mu)}
{a^{(1)}(\l-\mu)}B^{(1)}_c(\mu)A^{(1)}(\l)],
\no\\
B^{(1)}_{a_1}(\l)B^{(1)}_{a_2}(\mu) &=&
r^{(1)}(\l-\mu)^{b_1a_2}_{b_2a_1}B^{(1)}_{b_2}(\mu)B^{(1)}_{b_1}(\l).
\no \eea
Here the values 1, 2 are both  fermionic ($\e_1=1=\e_2$). The
$R$-matrix $r^{(1)}(\mu)$ is
\bea
{r^{(1)}(u)}^{ab}_{cd}&=&b^{(1)}(u){I^{(2)}}^{ab}_{cd} +a^{(1)}(u)
{\Pi^{(2)}}^{ab}_{cd}\no\\
a^{(1)}(u)&=&\frac{u}{1+s_5u},~~~b^{(1)}(u)=\frac{1}{1+s_5u}.
\no \eea
Above ${\Pi^{(2)}}^{ab}_{cd}$ is a $4\times4$ submatrix of $\Pi^{(1)}$ 
\bea
{\Pi}^{(2)}=
\left ( \begin {array} {cccc}
-s_4&0&0&0\\
0&0&-s_4&0\\
0&-s_4&0&0\\
0&0&0&-s_4
\end {array}  \right ),
\label{R2}
\eea
corresponding to the
grading $\e_1=\e_2=1$.  
As the reference state for the first nesting we choose 
${|0\re\,}^{(1)}_k=(1,0,0)^t, {|0\re\,}^{(1)}
=\otimes ^n_{k=1}{ |0\re\,}^{(1)}_k$ as
the pseudo-vacuum. We find
\bea
A^{(1)}(u) {|0\re\,}^{(1)} &=& 
 {|0\re\,}^{(1)} ,\no\\
D^{(1)}_{11}(u) {|0\re\,}^{(1)} &=& 
D^{(1)}_{22}(u) {|0\re\,}^{(1)}=\prod^n_{j=1}s_2 a^{(1)}(u-\l_j) 
 {|0\re\,}^{(1)},
\no \eea
and due to $\tau^{(1)}(u)=\frac{1}{s_3} A^{(1)}(u)+D^{(1)}_{11}(u)
+D^{(1)}_{22}(u)$ we get the eigenvalue
$$
\Lambda^{(1)}(u,\{\l_k\})
=\frac{1}{s_3}\prod^{n_1}_{j=1}\frac{1}{s_2a^{(1)}(\l^{(1)}_j-u)}
+\prod^{n_1}_{j=1}\frac{1}{s_2a^{(1)}(u-\l^{(1)}_j)}
\prod^{n}_{k=1}s_2a^{(1)}(u-\l_k)
\Lambda^{(2)}(u,\{\l^{(1)}_m\}),
$$
provided the parameters $\{\l^{(1)}_m\}$ satisfy
$$
\Lambda^{(2)}(\l^{(1)}_m)
=\frac{1}{s_3}
\prod^{n_1}_{\stackrel{l=1}{l \neq m}}\frac{a^{(1)}(\l^{(1)}_m-\l^{(1)}_l)}
{a^{(1)}(\l^{(1)}_l-\l^{(1)}_m)}
\prod^{n}_{k=1}\frac{1}{s_2a^{(1)}(\l^{(1)}_m-\l_k)}.$$
The transfer matrix of the second nesting is written as
$$
\tau^{(2)}(u,\{\l^{(1)}_m\})=str[{L}^{(2)}_{n_1}(u-\l^{(1)}_{n_1})
{L}^{(2)}_{n_1-1}(u-\l^{(1)}_{n_1-1}) \cdots {L}^{(2)}_1(u-\l^{(1)}_1)],
$$
where
\beq
{L}^{(2)}_k(u)=
 \left ( \begin {array}
{cc}
 a^{(2)}(u)-b^{(2)}(u)e^{11}_k&-b^{(2)}(u)e^{21}_k\\
 -b^{(2)}(u)e^{12}_k&a^{(2)}-b^{(2)}(u)e^{22}_k
     \end {array} \right ). 
\label{T211}
\eeq
From the Yang-Baxter relation for $\tau^{(2)}(u)$ one can derive the
following intertwining relation
\bea
r^{(1)}(\l-\mu){T}^{(2)}_{n_1}(\l)\otimes {T}^{(2)}_{n_1}(\mu)
={T}^{(2)}_{n_1}(\mu)\otimes {T}^{(2)}_{n_1}(\l) r^{(1)}(\l-\mu).
\label{YBAr1}
\eea
The components of (\ref{YBAr1}) needed for the construction of an algebraic
Bethe ansatz are
\bea
D^{(2)}(\mu)B^{(2)}(\l) &=&\frac{1}{s_4}[
\frac{1}{a^{(2)}(\l-\mu)} B^{(2)}(\l)D^{(2)}(\mu)
+\frac{b^{(2)}(\mu-\l)}{a^{(2)}(\mu-\l)}B^{(2)}(\mu)D^{(2)}(\l)],\no\\
A^{(2)}(\mu)B^{(2)}(\l) &=&\frac{1}{s_4}[
\frac{1}{a^{(2)}(\mu-\l)} B^{(2)}(\l)A^{(2)}(\mu)
+\frac{b^{(2)}(\l-\mu)}{a^{(2)}(\l-\mu)}B^{(2)}(\mu)A^{(2)}(\l)],\no\\
B^{(2)}(\l)B^{(2)}(\mu) &=&
B^{(2)}(\mu)B^{(2)}(\l), 
\label{comm} \eea
where
\bea
a^{(2)}(u)&=&\frac{u}{1+s_4u},~~~b^{(2)}(u)=\frac{1}{1+s_4u}.
\no \eea
For the reference state for the second nesting we pick 
${|0\re\,}^{(2)}_k=(1,0)^t, {|0\re\,}^{(2)}
=\otimes ^{n_1}_{k=1}{ |0\re\,}^{(2)}_k$. From the action of the nested
monodromy matrix
$$
{T}^{(2)}_{n_1}(u)={L}^{(2)}_{n_1}(u){L}^{(2)}_{n_1-1}(u)
\cdots {L}^{(2)}_1(u) \equiv
 \left ( \begin {array}
{cc}
 A^{(2)}(u)&B^{(2)}(u)\\
 C^{(2)}(u)& D^{(2)}(u)
     \end {array} \right ),
$$
we find 
$$
A^{(2)}(u) {|0\re\,}^{(2)}=\prod^{n_1}_{j=1}
\frac{a^{(2)}(u-\l^{(1)}_j)}
{a^{(2)}(\l^{(1)}_j-u)}
 {|0\re\,}^{(1)} ,~~~~
D^{(2)}(u) {|0\re\,}^{(1)} = 
\prod^{n_1}_{j=1}s_4 a^{(2)}(u-\l^{(1)}_j) 
 {|0\re\,}^{(1)} ,
$$
due to $\tau^{(2)}(u)= -A^{(2)}(u)-D^{(2)}(u) $. Thus 
$$
\Lambda^{(2)}(u,\{\l^{(1)}_m\})
=-[\prod^{n_2}_{j=1}\frac{1}{s_4a^{(2)}(u-\l^{(2)}_j)}
\prod^{n_1}_{m=1}\frac {a^{(2)}(u-\l^{(1)}_m)}
{a^{(2)}(\l^{(1)}_m-u)}
+\prod^{n_2}_{j=1}\frac{1}{s_4a^{(2)}(\l^{(2)}_j-u)}
\prod^{n_1}_{m=1}s_4a^{(2)}(u-\l^{(1)}_m)
$$
under the condition that the spectral parameters $\{\l^{(2)}_p\}$ are
solutions to the Bethe ansatz equation
$$
\prod^{n_2}_{\stackrel{j=1}{j \neq p}}\frac{a^{(2)}(\l^{(2)}_j-\l^{(2)}_p)}
{a^{(2)}(\l^{(2)}_p-\l^{(2)}_j)}
=\prod^{n_1}_{k=1}s_4a^{(2)}(\l^{(1)}_k-\l^{(2)}_p).
$$

We have now obtained the complete set of nested Bethe ansatz equations,
which read 
\bea
\left(\frac{1+s_1\l_k}{\l_k}\right)^L&=&s_3^{N_l-1}(-1)^{-N_e}
s_3^{L-n}(-1)^{N_e+N_{\downarrow}}\no\\
&&\times
\prod^{N_e}_{j=1}\frac{s_5(\l^{(1)}_j-\l_k)+1}
{s_2(\l^{(1)}_j-\l_k)}
\prod^{N_e+N_l}_{\stackrel{l=1}{l \neq k}}\frac{s_1(\l_k-\l_l)+1}
{s_1(\l_k-\l_l)-1},\no\\
\prod^{N_e+N_l}_{j=1}\frac{s_5(\l^{(1)}_m-\l_k)+1}
{s_2(\l^{(1)}_m-\l_k)}&=&s_3
\prod^{N_e}_{\stackrel{l=1}{l \neq m}}\frac{s_5(\l^{(1)}_m-\l^{(1)}_l)+1}
{s_5(\l^{(1)}_m-\l^{(1)}_l)-1}
\frac{s_4(\l^{(1)}_m-\l^{(1)}_l)-1}
{s_4(\l^{(1)}_m-\l^{(1)}_l)+1}
\prod^{N_{\downarrow}}_{j=1} \frac{s_4(\l^{(1)}_m-\l^{(2)}_j)+1}
{s_4(\l^{(1)}_m-\l^{(2)}_j)},\no\\
\prod^{N_{\downarrow}}_{\stackrel{j=1}{j \neq p}}
\frac{s_4(\l^{(2)}_j-\l^{(2)}_p)-1}
{s_4(\l^{(2)}_j-\l^{(2)}_p)+1}&=&
\prod^{N_e}_{k=1} \frac{s_4(\l^{(1)}_k-\l^{(2)}_p)}
{s_4(\l^{(1)}_k-\l^{(2)}_p)+1}.
\no \eea
Here we have used $n=N_e+N_l=N_{\downarrow}+N_{\uparrow}+N_l,n_1=N_e,
n_2=N_{\downarrow}$. 
The corresponding energy eigenvalue $E$ of the model
is given by
\beq
       E=\sum ^{N_e+N_l}_{j=1} \frac {1}{\l_j(1+s_1 \l_j)}-L.
\label{E}
       \eeq

The BBFF grading solution of above models closely follows the solution
of the EKS model in \cite{EKS1,Sch,EK1}.

\section{algebraic Bethe ansatz for group 2 }

The algebraic Bethe ansatz calculations for this group proceed in
exactly the same manner as group 1 up to the introduction of the matrix
$\Pi^{(1)}$, which for this case reads
\bea
{\Pi}^{(1)}=
\left ( \begin {array} {ccccccccccc}
s_5&0&0&0&0&0&0&0&0\\
0&0&0&s_2&0&0&0&0&0\\
0&0&0&0&0&0&s_2&0&0\\
0&s_2&0&0&0&0&0&0&0\\
0&0&0&0&s_4&0&0&0&0\\
0&0&0&0&0&s_4&0&0&0\\
0&0&s_2&0&0&0&0&0&0\\
0&0&0&0&0&0&0&s_4&0\\
0&0&0&0&0&0&0&0&s_4
\end {array}  \right ). 
\no
\eea
Following the calculation along the same lines as the previous section,
we find that the matrix $r^{(1)}(u)$ appearing in the equations (\ref{comm})
is of the form 
$$
{r^{(1)}(u)}^{bb}_{aa}=(b^{(1)}(u)+s_4a^{(1)}(u)){I^{(2)}}^{bb}_{aa}
=\frac{1+s_4u}{1+s_5u}{I^{(2)}}^{bb}_{aa}. 
$$
Here ${I^{(2)}}^{bb}_{aa}$ is the $4\times4$ identity matrix.  
For the reference state of the first nesting we choose the state
${|0\re\,}^{(1)}_k=(1,0,0)^t, {|0\re\,}^{(1)}
=\otimes ^n_{k=1}{ |0\re\,}^{(1)}_k$ as
the pseudo-vacuum and then find
\bea
A^{(1)}(u) {|0\re\,}^{(1)} &=& 
 {|0\re\,}^{(1)} ,\no\\
D^{(1)}_{11}(u) {|0\re\,}^{(1)} &=& 
D^{(1)}_{22}(u) {|0\re\,}^{(1)}=\prod^n_{j=1}s_2 a^{(1)}(u-\l_j) 
 {|0\re\,}^{(1)}. 
\no \eea
Due to $\tau^{(1)}(u)=\frac{1}{s_3} A^{(1)}(u)+D^{(1)}_{11}(u)
+D^{(1)}_{22}(u)$, we get the eigenvalue
$$
\Lambda^{(1)}(u,\{\l_k\})
=\frac{1}{s_3}\prod^{n_1}_{j=1}\frac{1}{s_2a^{(1)}(\l^{(1)}_j-u)}
+\prod^{n_1}_{j=1}\frac{1}{s_2a^{(1)}(u-\l^{(1)}_j)}
\prod^{n}_{k=1}s_2a^{(1)}(u-\l_k)
$$
under the condition that the spectral parameters $\{\l^{(1)}_m\}$ are
solutions of the Bethe ansatz equation
$$
\prod^{n_1}_{\stackrel{l=1}{l \neq m}}\frac{a^{(1)}(\l^{(1)}_m-\l^{(1)}_l)}
{a^{(1)}(\l^{(1)}_l-\l^{(1)}_m)}=s_3
\prod^{n}_{k=1}s_2a^{(1)}(\l^{(1)}_m-\l_k).$$
For the full solution, we have the nested Bethe ansatz equations 
\bea
\left(\frac{1+s_1\l_k}{\l_k}\right)^L&=&s_3^{N_l-1}(-1)^{-N_e}
s_3^{L-n}(-1)^{N_e+N_{\downarrow}}\no\\
&&\times\prod^{N_e}_{j=1}\frac{s_5(\l^{(1)}_j-\l_k)+1}
{s_2(\l^{(1)}_j-\l_k)}
\prod^{N_e+N_l}_{\stackrel{l=1}{l \neq k}}\frac{s_1(\l_k-\l_l)+1}
{s_1(\l_k-\l_l)-1},\no\\
\prod^{N_e+N_l}_{j=1}\frac{s_5(\l^{(1)}_m-\l_k)+1}
{s_2(\l^{(1)}_m-\l_k)}&=&\frac{1}{s_3}
\prod^{N_e}_{\stackrel{l=1}{l \neq m}}\frac{s_5(\l^{(1)}_m-\l^{(1)}_l)+1}
{s_5(\l^{(1)}_m-\l^{(1)}_l)-1},
\no \eea
where $n,\,n_1,\,n_2$ have the same meaning as previously. Also, the
energy expression (\ref{E}) applies here.

\section{algebraic Bethe ansatz for group 3}

We now consider the case of the algebraic Bethe ansatz
for group 3. As we will see here,
the procedure is fundamentally different from the preceding cases in
that we are required to work with a subspace of reference states for the
first level of the algebraic Bethe ansatz.
The methodology we employ follows that proposed
by Abad and R\'{\i}os \cite{AR}.

In the case of group 3, the $R$-matrix reads 
\small
\bea
&&\check{R}(u)=1+u\Pi=\no\\
&&\left ( \begin {array} {cccccccccccccccc}
1+s_1u&0&0&0&0&0&0&0&0&0&0&0&0&0&0&0\\
0&1+s_1u&0&0&0&0&0&0&0&0&0&0&0&0&0&0\\
0&0&1&0&0&0&0&0&-u&0&0&0&0&0&0&0\\
0&0&0&1&0&0&0&0&0&0&0&0&-u&0&0&0\\
0&0&0&0&1+s_1u&0&0&0&0&0&0&0&0&0&0&0\\
0&0&0&0&0&1+s_1u&0&0&0&0&0&0&0&0&0&0\\
0&0&0&0&0&0&1&0&0&s_2u&0&0&0&0&0&0\\
0&0&0&0&0&0&0&1&0&0&0&0&0&s_2u&0&0\\
0&0&-u&0&0&0&0&0&1&0&0&0&0&0&0&0\\
0&0&0&0&0&0&s_2u&0&0&1&0&0&0&0&0&0\\
0&0&0&0&0&0&0&0&0&0&1-s_3u&0&0&0&0&0\\
0&0&0&0&0&0&0&0&0&0&0&1&0&0&-s_3u&0\\
0&0&0&-u&0&0&0&0&0&0&0&0&1&0&0&0\\
0&0&0&0&0&0&0&s_2u&0&0&0&0&0&1&0&0\\
0&0&0&0&0&0&0&0&0&0&0&-s_3u&0&0&1&0\\
0&0&0&0&0&0&0&0&0&0&0&0&0&0&0&1-s_3u
\end {array}  \right ),
\label{R3}
\eea
\normalsize
and we again express the $L$ operator as 
$$
{L}_j(u)=\frac{1}{1+s_1u}P\check{R}(u). 
$$
If we choose the local vacuum state as $|0\re\,_j=\frac{1}{\sqrt {\a^2+\b^2}}
(\a,\b,0,0)^t$, and act
the $L$-operator on this local vacuum state, we have
\beq
{L}_j(u){|0\re\,}_j=
 \left ( \begin {array}
{cccc}
 e_{11}&e_{21}&*&*\\
 e_{12}& e_{22}&*&*\\
 0&0&(s_2e_{22}-e_{11})a(u)&0\\ 
 0&0&0&(s_2e_{22}-e_{11})a(u) 
     \end {array} \right ){|0\re\,}_j.
\label{u3}
\eeq
Define the vacuum state as $|0\re\,=\otimes^L_{j=1}{|0\re\,}_j$
and represent the monodromy matrix as
\beq
{T}(u)={L}_L(u){L}_{L-1}(u)\cdots {L}_1(u) \equiv
 \left ( \begin {array}
{cccc}
 A_{11}(u)&A_{12}(u)& B_{11}(u)&B_{12}(u)\\
 A_{21}(u)&A_{22}(u)& B_{21}(u)&B_{22}(u)\\
 C_{11}(u)&C_{12}(u)& D_{11}(u)&D_{12}(u)\\
 C_{21}(u)&C_{22}(u)& D_{21}(u)&D_{22}(u)
     \end {array} \right ). 
\label{T13}
\eeq
The transfer matrix is thus written explicitly as 
$$
\tau(u)=A_{11}(u)+A_{22}(u)-D_{11}(u)-D_{22}(u).
$$
The action of the monodromy matrix on the vacuum state is 
\bea
[A_{11}(u)+A_{22}(u)]|0\re\,&=& tr_0[P_{L0}P_{L-1,0}\cdots P_{10}]|0\re\,
\no\\
D_{11}(u)|0\re\,&=&D_{22}(u)|0\re\,=[a(u)]^L(-1)^{N_e} s_2^{N_l},\no\\
B_{ik}(u)|0\re\,&\neq& 0,~~~~~C_{ik}(u)|0\re\,=0~~~~~~~(i,k=1,2),
\no \eea
where $P_{j0}$ is the permutation operator for two-dimensional spaces
(corresponding to the indices 1 and 2). 
Substituting (\ref{T13}) into the Yang-Baxter algebra (\ref {YBA}) ,
we may deduce the following commutation relations:
\bea
D_{ac}(\mu)B_{bd}(\l) &=&
S_b[(-1)^{\e_a\e_b}\frac{r(\mu-\l)^{d'd}_{c'c}}{a(\mu-\l)}B_{ac'}(\l)
D_{bd'}(\mu)
-(-1)^{\e_a\e_b}\frac{b(\mu-\l)}{a(\mu-\l)}B_{bc}(\mu)D_{ad}(\l)],\no\\
A_{ac}(\mu)B_{bd}(\l) &=&
S_c[\frac{1}{a(\l-\mu)}B_{ad}(\l)A_{bc}(\mu)-\frac{b(\l-\mu)}{a(\l-\mu)}
B_{ad}(\mu)A_{bc}(\l)], \no\\
B_{ac}(\l)B_{bd}(\mu) &=&
r(\l-\mu)^{d'd}_{c'c}B_{ac'}(\mu)B_{bd'}(\l),
\label{comm1} 
\eea
where
$$
{r(u)}^{ab}_{cd}=b(u){I^{(2)}}^{ab}_{cd}+a(u){\Pi^{(2)}}^{ab}_{cd}.
$$
Here ${\Pi^{(2)}}^{ab}_{cd}=s_3P^{(2)}$ with permutation matrix
$P^{(2)} =-\sum_{ij}e_{ij}\otimes e_{ji}$ 
corresponding to the grading $\e_1=\e_2=1$ and $S_1=-1,S_2=\frac{1}{s_2} $.
Denote the eigenvalues of $\tau(u)$ and $\tau^{(1)}$
by $\Lambda(u)$ and $\Lambda^{(1)}(u)$. We now have
$$
\Lambda(u)
=G \cdot \prod^{n_1}_{j=1}\frac{1}{a(\l_j-u)}
+[a(u)]^L \prod^{n_1}_{j=1}\frac{1}{a(u-\l_j)}\Lambda^{(1)}(u).
$$
Here $G= {\rm tr[diag}(-1,\frac{1}{s_2})P_{L0}P_{L-1,0}\cdots P_{10}]$,
and the parameters $\{\l_k\}$ are required to satisfy the
Bethe ansatz equations
$$
\Lambda^{(1)}(\l_k)
=[a(\l_k)]^{(-L)}\cdot G \cdot
\prod^{n_1}_{\stackrel{j=1}{j \neq k}}\frac{a(\l_k-\l_j)}{a(\l_j-\l_k)}. $$
The nested transfer matrix is written as the supertrace on the auxiliary
space for the reduced monodromy matrix which satisfies the Yang-Baxter
relation, {\it i.e.}
$$
\tau^{(1)}(u,\{\l_k\})=
{\rm str[diag}(-1,\frac{1}{s_2}){L}^{(1)}_{n_1}(u-\l_{n_1})
{L}^{(1)}_{n_1-1}(u-\l_{n_1-1}) \cdots {L}^{(1)}_1(u-\l_1)],
$$
\bea
r(\l-\mu){T}^{(1)}_{n_1}(\l)\otimes{T}^{(1)}_{n_1}(\mu)
={T}^{(1)}_{n_1}(\mu)\otimes {T}^{(1)}_{n_1}(\l) r(\l-\mu).
\label{YBAr3}
\eea
The components of (\ref{YBAr3}) needed for the construction of an algebraic
Bethe ansatz are
\bea
D^{(1)}(\mu)B^{(1)}(\l) &=&\frac{1}{s_3}[
\frac{1}{a^{(3)}(\l-\mu)} B^{(1)}(\l)D^{(1)}(\mu)
+\frac{b^{(3)}(\mu-\l)}{a^{(3)}(\mu-\l)}B^{(1)}(\mu)D^{(1)}(\l)],\no\\
A^{(1)}(\mu)B^{(1)}(\l) &=&\frac{1}{s_3}[
\frac{1}{a^{(3)}(\mu-\l)} B^{(1)}(\l)A^{(1)}(\mu)
+\frac{b^{(3)}(\l-\mu)}{a^{(3)}(\l-\mu)}B^{(1)}(\mu)A^{(1)}(\l)],\no\\
B^{(1)}(\l)B^{(1)}(\mu) &=&
B^{(1)}(\mu)B^{(1)}(\l), 
\no \eea
where
\bea
a^{(3)}(u)&=&\frac{u}{1+s_3u},~~~b^{(3)}(u)=\frac{1}{1+s_3u}.
\no \eea
As the reference state for the second nesting we take  
${|0\re\,}^{(1)}_k=(1,0)^t, {|0\re\,}^{(1)}
=\otimes ^{n_1}_{k=1}{ |0\re\,}^{(2)}_k$. From the action of the nested
monodromy matrix
$$
{T}^{(1)}_{n_1}(u)={L}^{(1)}_{n_1}(u){L}^{(1)}_{n_1-1}(u)
\cdots {L}^{(1)}_1(u) \equiv
 \left ( \begin {array}
{cc}
 A^{(1)}(u)&B^{(1)}(u)\\
 C^{(1)}(u)& D^{(1)}(u)
     \end {array} \right ),
$$
we find 
$$
A^{(1)}(u) {|0\re\,}^{(1)}=\prod^{n_1}_{j=1}
\frac{a^{(3)}(u-\l_j)}
{a^{(3)}(\l_j-u)}
 {|0\re\,}^{(1)} ,~~~~
D^{(1)}(u) {|0\re\,}^{(1)} = 
\prod^{n_1}_{j=1}s_3 a^{(3)}(u-\l_j) 
 {|0\re\,}^{(1)} ,
$$
and due to $\tau^{(1)}(u)=A^{(1)}(u)-\frac{1}{s_2}D^{(1)}(u)
$ we have
$$
\Lambda^{(1)}(u,\{\l_k\})
=\prod^{n_2}_{j=1}\frac{1}{s_3a^{(3)}(u-\l^{(1)}_j)}
\prod^{n_1}_{k=1}\frac {a^{(3)}(u-\l_k)}
{a^{(3)}(\l_k-u)}
-\frac{1}{s_2}\prod^{n_2}_{j=1}\frac{1}{s_3a^{(3)}(\l^{(1)}_j-u)}
\prod^{n_1}_{k=1}s_3a^{(3)}(u-\l_k)
$$
under the condition that the spectral parameters $\{\l^{(1)}_m\}$ are
solutions to the Bethe ansatz equation
$$
\prod^{n_2}_{\stackrel{j=1}{j \neq m}}\frac{a^{(3)}(\l^{(1)}_j-\l^{(1)}_m)}
{a^{(3)}(\l^{(1)}_m-\l^{(1)}_j)}
=-\frac{1}{s_2}\prod^{n_1}_{k=1}s_3a^{(3)}(\l_k-\l^{(1)}_m)
. $$ 
Now we obtain the complete set of nested Bethe ansatz equations reading  
\bea
\left(\frac{1+s_1\l_k}{\l_k}\right)^L&=&
-\frac{1}{G} \cdot
\prod^{N_e}_{\stackrel{j=1}{j \neq k}}\frac{s_1(\l_k-\l_j)-1}
{s_1(\l_k-\l_j)+1}
\frac{s_1(\l_k-\l_l)-1}
{s_1(\l_k-\l_l)+1}
\prod^{N_{\downarrow}}_{j=1}\frac{s_3(\l^{(1)}_j-\l_k)-1}
{s_3(\l^{(1)}_j-\l_k)},\no\\
\prod^{N_{\downarrow}}_{\stackrel{j=1}{j \neq p}}
\frac{s_3(\l^{(1)}_j-\l^{(1)}_p)-1}
{s_3(\l^{(1)}_j-\l^{(1)}_p)+1}&=&
-\frac{1}{s_2}\prod^{N_e}_{k=1} \frac{s_3(\l_k-\l^{(1)}_p)}
{s_3(\l_k-\l^{(1)}_p)+1}.
\no \eea
The energy expression for this model reads the same as in the previous
cases (\ref{E}).

\section{algebraic Bethe ansatz for group 4}

Just as the calculations for the cases of group 1 and 2 follow along similar
lines, we find an analogous situation occurring with groups 3 and 4.
For group 4, we have the $R$-matrix 
\small
\bea
&&\check{R}(u)=1+u\Pi=\no\\
&&\left ( \begin {array} {cccccccccccccccc}
1+s_1u&0&0&0&0&0&0&0&0&0&0&0&0&0&0&0\\
0&1+s_1u&0&0&0&0&0&0&0&0&0&0&0&0&0&0\\
0&0&1&0&0&0&0&0&-u&0&0&0&0&0&0&0\\
0&0&0&1&0&0&0&0&0&0&0&0&-u&0&0&0\\
0&0&0&0&1+s_1u&0&0&0&0&0&0&0&0&0&0&0\\
0&0&0&0&0&1+s_1u&0&0&0&0&0&0&0&0&0&0\\
0&0&0&0&0&0&1&0&0&s_2u&0&0&0&0&0&0\\
0&0&0&0&0&0&0&1&0&0&0&0&0&s_2u&0&0\\
0&0&-u&0&0&0&0&0&1&0&0&0&0&0&0&0\\
0&0&0&0&0&0&s_2u&0&0&1&0&0&0&0&0&0\\
0&0&0&0&0&0&0&0&0&0&1+s_3u&0&0&0&0&0\\
0&0&0&0&0&0&0&0&0&0&0&1+s_3u&0&0&0&0\\
0&0&0&-u&0&0&0&0&0&0&0&0&1&0&0&0\\
0&0&0&0&0&0&0&s_2u&0&0&0&0&0&1&0&0\\
0&0&0&0&0&0&0&0&0&0&0&0&0&0&1+s_3u&0\\
0&0&0&0&0&0&0&0&0&0&0&0&0&0&0&1+s_3u
\end {array}  \right ). 
\label{R4}
\eea
\normalsize
The calculations of the algebraic Bethe ansatz
proceed in exactly the same manner as the
group 3 case except now we find that in (\ref{comm1}) we have 
$$
{r(u)}^{bb}_{aa}=(b(u)+s_3a(u)){I^{(2)}}^{bb}_{aa}=
\frac{1+s_3u}{1+s_1u}{I^{(2)}}^{bb}_{aa}.
$$
For the eigenvalues $\Lambda(u)$ of $\tau(u)$ 
we obtain the expression  
$$
\Lambda(u)
=G \cdot \prod^{n_1}_{j=1}\frac{1}{a(\l_j-u)}
+[a(u)]^L \prod^{n_1}_{j=1}\frac{1}{a(u-\l_j)},
$$
where $G$ is defined as before  and 
 $\{\l_k\}$ are subject to the  Bethe
ansatz equations
$$
\left(\frac{1+s_1\l_k}{\l_k}\right)^L= G \cdot
\prod^{N_e}_{\stackrel{j=1}{j \neq k}}\frac{s_1(\l_k-\l_j)+1}
{s_1(\l_k-\l_j)-1}.
$$
Again, the energies are given by (\ref{E}).

\section{algebraic Bethe ansatz for group 5}

For group 5, the  $R$-matrix is given by  
\small
\bea
&&\check{R}(u)=1+u\Pi=\no\\
&&\left ( \begin {array} {cccccccccccccccc}
1+s_1u&0&0&0&0&0&0&0&0&0&0&0&0&0&0&0\\
0&1&0&0&s_2u&0&0&0&0&0&0&0&0&0&0&0\\
0&0&1&0&0&0&0&0&-u&0&0&0&0&0&0&0\\
0&0&0&1&0&0&0&0&0&0&0&0&-u&0&0&0\\
0&s_2u&0&0&1&0&0&0&0&0&0&0&0&0&0&0\\
0&0&0&0&0&1+s_3u&0&0&0&0&0&0&0&0&0&0\\
0&0&0&0&0&0&1+s_3u&0&0&0&0&0&0&0&0&0\\
0&0&0&0&0&0&0&1+s_3u&0&0&0&0&0&0&0&0\\
0&0&-u&0&0&0&0&0&1&0&0&0&0&0&0&0\\
0&0&0&0&0&0&0&0&0&1+s_3u&0&0&0&0&0&0\\
0&0&0&0&0&0&0&0&0&0&1+s_3u&0&0&0&0&0\\
0&0&0&0&0&0&0&0&0&0&0&1+s_3u&0&0&0&0\\
0&0&0&-u&0&0&0&0&0&0&0&0&1&0&0&0\\
0&0&0&0&0&0&0&0&0&0&0&0&0&1+s_3u&0&0\\
0&0&0&0&0&0&0&0&0&0&0&0&0&0&1+s_3u&0\\
0&0&0&0&0&0&0&0&0&0&0&0&0&0&0&1+s_3u
\end {array}  \right ).
\label{R5}
\eea
\normalsize
We choose the local vacuum state as $|0\re\,_j=(1,0,0,0)^t$. Acting
the $L$-operator on this local vacuum state, we have
\beq
{L}_j(u){|0\re\,}_j=
 \left ( \begin {array}
{cccc}
 1&*&*&*\\
 0& s_2a(u)&0&0\\
 0&0&-a(u)&0\\ 
 0&0&0&-a(u)
     \end {array} \right ){|0\re\,}_j.
\label{u5}
\eeq
Define the vacuum state as $|0\re\,=\otimes^L_{j=1}{|0\re\,}_j$. 
The monodromy matrix is represented as
\beq
{T}(u)={L}_L(u){L}_{L-1}(u)\cdots {L}_1(u) \equiv
 \left ( \begin {array}
{cccc}
 A(u)&B_1(u)& B_2(u)&B_2(u)\\
 C_1(u)& D_{11}(u)&D_{12}(u)&D_{13}(u)\\
 C_2(u)&D_{21} (u)& D_{22}(u)&D_{23}(u)\\
 C_3(u)&D_{31} (u)& D_{32}(u)&D_{33}(u)
     \end {array} \right ),
\label{T15}
\eeq
and so the  transfer matrix is explicitly  
$$
\tau(u)=A(u)+D_{11}(u)-D_{22}(u)-D_{33}(u).
$$
The action of the monodromy matrix on the vacuum state is 
\bea
A(u)|0\re\,&=&|0\re\,,~~~~~D_{11}(u)=[s_2a(u)]^L|0\re\,,
~~~D_{22}(u)=D_{33}(u)=[-a(u)]^L|0\re\,\no\\
B_k(u)|0\re\,& \neq &0,~~~~~C_k(u)=0,~~~~~D_{ik}(u)=0,~~~~~~~~( i \neq
k,~~i,k=1,2,3).
\label{T151}
\eea
Substituting (\ref{T15}) into the Yang-Baxter algebra (\ref {YBA}) ,
we may deduce the following commutation relations:
\bea
D_{ab}(\mu)B_c(\l) &=&
S_a[(-1)^{\e_a\e_b}\frac{r(\mu-\l)^{bb}_{aa}}{a(\mu-\l)}B_b(\l)D_{ac}(\mu)
-(-1)^{\e_a\e_b}\frac{b(\mu-\l)}{a(\mu-\l)}B_b(\mu)D_{ac}(\l)],\no\\
A(\mu)B_c(\l) &=&
S_c[\frac{1}{a(\l-\mu)}B_c(\l)A(\mu)-\frac{b(\l-\mu)}{a(\l-\mu)}B_c(\mu)A(\l)],
\no\\
B_{a_1}(\l)B_{a_2}(\mu) &=&
r(\l-\mu)^{a_2a_2}_{a_1a_1}B_{a_1}(\mu)B_{a_2}(\l),
\no \eea
where
$$
{r(u)}^{bb}_{aa}=(b(u)+s_3a(u)){I^{(1)}}^{bb}_{aa}
=\frac{1+s_3u}{1+s_1u}{I^{(1)}}^{bb}_{aa}.  
$$
Here ${I^{(1)}}^{bb}_{aa}$ is the $9\times9$ id matrix
and $S_1=\frac{1}{s_2} 
,S_2=-1, S_3=-1$. The eigenvalues of $\tau(u)$ read
$$
\Lambda(u)
=s_2^{-N_l}(-1)^{N_e}\prod^n_{j=1}\frac{1}{a(\l_j-u)}
+[a(u)]^Ls_2^{L-n}(-1)^{N_e+N_{\downarrow}}
\prod^n_{j=1}\frac{1}{a(u-\l_j)}
$$
with the following Bethe
ansatz equations 
$$
\left(\frac{1+s_1\l_k}{\l_k}\right)^L=s_2^{N_l}(-1)^{-N_e}
s_2^{L-n}(-1)^{N_e+N_{\downarrow}}
\prod^{N_e+N_l}_{\stackrel{k=1}{k \neq j}}\frac{s_1(\l_j-\l_k)+1}
{s_1(\l_j-\l_k)-1} $$ 
and the energy is given by (\ref{E}). 

\section{algebraic Bethe ansatz for group 6}

The final case to consider corresponds to the 
$R$-matrix  
\small
\bea
&&\check{R}(u)=1+u\Pi=\no\\
&&\left ( \begin {array} {cccccccccccccccc}
1+s_1u&0&0&0&0&0&0&0&0&0&0&0&0&0&0&0\\
0&1&0&0&s_2u&0&0&0&0&0&0&0&0&0&0&0\\
0&0&1+s_1u&0&0&0&0&0&0&0&0&0&0&0&0&0\\
0&0&0&1+s_1u&0&0&0&0&0&0&0&0&0&0&0&0\\
0&s_2u&0&0&1&0&0&0&0&0&0&0&0&0&0&0\\
0&0&0&0&0&1+s_1u&0&0&0&0&0&0&0&0&0&0\\
0&0&0&0&0&0&1&0&0&u&0&0&0&0&0&0\\
0&0&0&0&0&0&0&1&0&0&0&0&0&u&0&0\\
0&0&0&0&0&0&0&0&1+s_1u&0&0&0&0&0&0&0\\
0&0&0&0&0&0&u&0&0&1&0&0&0&0&0&0\\
0&0&0&0&0&0&0&0&0&0&1+s_1u&0&0&0&0&0\\
0&0&0&0&0&0&0&0&0&0&0&1+s_1u&0&0&0&0\\
0&0&0&-u&0&0&0&0&0&0&0&0&1+s_1u&0&0&0\\
0&0&0&0&0&0&0&u&0&0&0&0&0&1&0&0\\
0&0&0&0&0&0&0&0&0&0&0&0&0&0&1+s_1u&0\\
0&0&0&0&0&0&0&0&0&0&0&0&0&0&0&1+s_1u
\end {array}  \right ).
\label{R6}
\eea
\normalsize
In contrast to the other cases considered,
we choose the local vacuum state as $|0\re\,_j=(0,1,0,0)^t$. Acting
the $L$-operator on this local vacuum state, we have
\beq
{L}_j(u){|0\re\,}_j=
 \left ( \begin {array}
{cccc}
 s_2a(u)&0&0&0\\
 \, *&1&*&*\\
 0&0&a(u)&0\\ 
 0&0&0&a(u)
     \end {array} \right ){|0\re\,}_j.
\label{u6}
\eeq
Defining the vacuum state as $|0\re\,=\otimes^L_{j=1}{|0\re\,}_j$
we express the monodromy matrix as
\beq
{T}(u)={L}_L(u){L}_{L-1}(u)\cdots {L}_1(u) \equiv
 \left ( \begin {array}
{cccc}
 D_{11}(u)&C_1(u)&D_{12}(u)&D_{13}(u)\\
 B_1(u)&A(u)& B_2(u)&B_2(u)\\
 D_{21} (u)&C_2(u)& D_{22}(u)&D_{23}(u)\\
 D_{31} (u)&C_3(u)& D_{32}(u)&D_{33}(u)
     \end {array} \right ),
\label{T16}
\eeq
and so the transfer matrix is
$$
\tau(u)=D_{11}(u)+A(u)-D_{22}(u)-D_{33}(u).
$$
The action of the monodromy matrix on the vacuum state is given by  
\bea
D_{11}(u)&=&[s_2a(u)]^L|0\re\,,~~~~~~
A(u)|0\re\,=|0\re\,,
~~~D_{22}(u)=D_{33}(u)=[a(u)]^L|0\re\,\no\\
B_k(u)|0\re\,& \neq &0,~~~~~C_k(u)=0,~~~~~D_{ik}(u)=0,~~~~~~~~( i \neq
k,~~i,k=1,2,3).
\label{T161}
\eea
Substituting (\ref{T16}) into the Yang-Baxter algebra (\ref {YBA})
we find
\bea
D_{ab}(\mu)B_c(\l) &=&
S_a[(-1)^{\e_a\e_b}\frac{r(\mu-\l)^{bb}_{aa}}{a^{(3)}(\mu-\l)}
B_b(\l)D_{ac}(\mu)
-(-1)^{\e_a\e_b}\frac{b^{(3)}(\mu-\l)}{a^{(3)}(\mu-\l)}B_b(\mu)D_{ac}(\l)],\no\\
A(\mu)B_c(\l) &=&
S_c[\frac{1}{a^{(3)}(\l-\mu)}B_c(\l)A(\mu)
-\frac{b^{(3)}(\l-\mu)}{a^{(3)}(\l-\mu)}B_c(\mu)A(\l)],
\no\\
B_{a_1}(\l)B_{a_2}(\mu) &=&
r(\l-\mu)^{a_2a_2}_{a_1a_1}B_{a_1}(\mu)B_{a_2}(\l),
\no \eea
where
$$
{r(u)}^{bb}_{aa}=(b^{(3)}(u)+s_1a^{(3)}(u)) {I^{(1)}}^{bb}_{aa} 
=\frac{1+s_1u}{1+s_3u}{I^{(1)}}^{bb}_{aa}  
$$
and now $S_1=\frac{1}{s_2},S_2=1,
S_3=1$. The eigenvalues for the transfer matrix read
$$
\Lambda(u)
=s_2^{-N_{\uparrow}}\prod^{L-N_l}_{j=1}\frac{1}{a^{(3)}(\l_j-u)}
+[a(u)]^Ls_2^{L-n}
\prod^{L-N_l}_{j=1}\frac{1}{a^{(3)}(u-\l_j)}
$$
so that the Bethe ansatz equations
$$
\left(\frac{1+s_1\l_k}{\l_k}\right)^L=s_2^{N_{\uparrow}}
s_2^{L-n}
\prod^{L-N_l}_{\stackrel{k=1}{k \neq j}}\frac{s_3(\l_j-\l_k)+1}
{s_3(\l_j-\l_k)-1}
$$
are satisfied. 
Here the energy eigenvalue differs somewhat from the previous cases and
has the form
$$ 
       E=\sum ^{L-N_l}_{j=1} \frac {1}{\l_j(1+s_1 \l_j)}-L.
$$

\section{summary and discussion}

In this paper, integrable extensions of the Hubbard model
arising from supersymmetric group solutions, by means of the algebraic
Bethe ansatz method, have been investigated. In particular, we have
calculated explicitly the Bethe ansatz equations as well as the energy
eigenvalues for 6 different classes of underlying $R$-matrices, which
in fact correspond to 96 different possible physical Hamiltonians.

A natural direction for possible further research is
to deal with physical applications of the above models. More specific
future works will be:
(i) studying low energy behaviour and physical properties of the
corresponding systems based on an analysis of the Bethe ansatz
equations from these results, including: investigating
the ground state structure, computing the finite size corrections to
the low-lying energies, and calculating thermodynamic equilibrium
properties, using the methods of Woynarovich \cite{Wo};
(ii) employing some traditional mathematical methods such as the
Wiener-Hopf technique to
solve the special kind of integral equations arising from the
thermodynamic Bethe ansatz equations, using
the methods of Yang \& Yang \cite{yang} and Babujian \cite{Bab}.

\vskip.3in
This work is supported by the Australian Research Council. 


\end{document}